\documentclass[aps,superscriptaddress,preprintnumbers,showpacs,twocolumn]{revtex4}

\usepackage{amssymb}

\usepackage{indentfirst}

\usepackage{epsfig}

\usepackage{epsf}

\usepackage{graphicx}

\def\Vec#1{{\bf #1}}

\def\D{{\mathrm d}}

\newcommand{\be}{\begin{equation}}
\newcommand{\ee}{\end{equation}}
\newcommand{\ba}{\begin{eqnarray}}
\newcommand{\ea}{\end{eqnarray}}

\newcommand{\p}{\perp}

\begin{document}
\newcommand*{\pku}{School of Physics and State Key Laboratory of Nuclear Physics and
Technology, Peking University, Beijing 100871,
China}\affiliation{\pku}
\newcommand*{\usm}{Departamento de F\'\i sica y Centro de Estudios
Subat\'omicos, Universidad T\'ecnica Federico Santa Mar\'\i a,
Casilla 110-V, Valpara\'\i so, Chile}\affiliation{\usm}

\title{$\cos 2 \phi$ asymmetries in unpolarized semi-inclusive DIS  }

\author{Bing Zhang}\affiliation{\pku}
\author{Zhun Lu}\affiliation{\usm}
\author{Bo-Qiang Ma}\email[Corresponding author. Electronic address: ]{mabq@phy.pku.edu.cn}\affiliation{\pku}
\author{Ivan Schmidt}\email[Corresponding author. Electronic address: ]{ivan.schmidt@usm.cl}\affiliation{\usm}

\begin{abstract}\
  We use the Boer-Mulders functions parameterized from unpolarized $p+D$ Drell-Yan data
by the FNAL E866/NuSea Collaboration combined with recently
extracted Collins functions to calculate the $\cos 2 \phi$
asymmetries in unpolarized semi-inclusive deeply inelastic
scattering (SIDIS) processes both for ZEUS at Hadron Electron Ring
Accelerator (HERA) and Jefferson Lab (JLab) experiments, and to
compare our results with their data. We also give predictions for
the $\cos 2 \phi$ asymmetries of SIDIS in the kinematical regime of
HERMES Collaboration, and the forthcoming JLab experiments. We
predict that the $\cos 2 \phi$ asymmetries of semi-inclusive $\pi^-$
production are somewhat larger than that of $\pi^+$ production. We
suggest to measure these two processes separately, which will
provide more detail information on the Boer-Mulders functions as
well as on the Collins functions.

\end{abstract}

\pacs{13.60.-r, 13.85.Ni, 13.88.+e, 14.20.Dh}

\maketitle


\section{Introduction}

The importance of the transverse-momentum-dependent distributions of
quarks for a full understanding of the structure of hadrons has been
widely recognized in the last decade
\cite{levelt,kotzinian,mulders,bm}. Experimentally semi-inclusive
deep inelastic scattering (SIDIS) provides a unique playground for
these $\Vec k_T$-dependent distributions, where the observables of
most interest are the single-spin asymmetries (SSA) and other
related asymmetries, which have been measured and are currently
under direct experimental scrutiny
\cite{Arneodo:1986cf,Air2004tw,hermes,
compass,compass07,belle06,zeus2000,zeus2006,jlab07}.

The leading-twist distributions, the Sivers function
$f_{1T}^{\perp}(x, \Vec k_T^2)$ \cite{sivers} and its chiral-odd
partner $h_1^{\perp}(x, \Vec k_T^2)$, the so-called Boer--Mulders
function~\cite{bm}, are greatly relevant to these asymmetries. These
two distributions describe the time-reversal odd correlations
between the intrinsic transverse momenta of quarks and transverse
spin vectors~\cite{bdr}. In particular, $f_{1T}^{\perp}$ represents
the distribution of unpolarized quarks inside a transversely
polarized hadron, whereas $h_1^{\perp}$ describes the transverse
spin distribution of quarks inside an unpolarized hadron.

The Sivers function is known to be responsible for the $\sin (\phi -
\phi_S)$
 single-spin asymmetry in transversely polarized SIDIS
\cite{Air2004tw,compass,compass07}, and has been extracted from data
by several
groups~\cite{anselmino05a,anselmino05b,efr05,cegmms,vy05,arnold08,anselmino08s}.
The Boer-Mulders function produces azimuthal asymmetries in
unpolarized reactions. Boer argued that it can account~\cite{boer}
for the observed $ \cos 2 \phi$ asymmetries in unpolarized $\pi N$
Drell-Yan processes \cite{na10,conway}. This is quantitatively
confirmed in \cite{lm04,lm05}, where it is shown to explain the
Drell-Yan dilepton asymmetry fairly well. Many other theoretical
calculations and phenomenological
analysis~\cite{gg02,bbh03,pobylista,yuan,bsy04,radici05,sissakian05,gamberg072,sissakian06,
lms06,lu06,lms07} have been performed on $h_1^{\p}$. Recently
lattice calculations ~\cite{lattice}, approaches based on
generalized parton distributions (GPD) ~\cite{gpds,burkardt07}, the
calculation in Ref.~\cite{gamberg07}, and also a new
quark-spectator-diquark calculation~\cite{bct08}
suggest that the $h_1^{\p}$ for $u$ and $d$ quarks are of the same
sign and of similar size. Apart from those result there is also
calculation from axial-diquark model calculation ~\cite{bsy04}
predicting that the $h_1^{\p}$ for $u$ and $d$ quarks are of
different sign. Experimentally the first measurement~\cite{e866} on
the $ \cos 2 \phi$ asymmetries in unpolarized Drell-Yan process with
proton beam has been performed by E866/NuSea Collaboration, which
provides further constraints on the Boer-Mulders functions of
nucleons. With the unpolarized $p+D$ Drell-Yan data available, first
attempt on extracting Boer-Mulders functions has been performed in
Ref.~\cite{bing08}.

Another phenomenological implication of $h_1^{\p}$ is its
consequences on the $\cos 2 \phi$ asymmetry in SIDIS, where $\phi$
is the azimuthal angle of the produced hadron related to the lepton
plane, as shown in Fig.~\ref{plane}. In order to generate the $\cos
2 \phi$ asymmetry that occurs in unpolarized SIDIS, there are three
possible mechanisms. 1) non-collinear kinematics at order
$k_T^2/Q^2$, the so-called Cahn effect \cite{cahn}; 2) the
leading-twist Boer-Mulders function coupling to a specular
fragmentation function, the so-called Collins function
\cite{collins93}, which describes the fragmentation of transversely
polarized quarks into unpolarized hadrons. The authors of
Refs.~\cite{vy05,egp06,anselmino07} extracted the favored and
unfavored Collins functions and suggested that the favored Collins
functions and the unfavored ones have opposite signs with comparable
absolute values; 3) perturbative gluon radiation
\cite{georgi,mendez,konig,chay}. In this paper we will employ the
first two mechanisms to study the $ \cos 2 \phi$ asymmetry in SIDIS
process. In the calculations we will adopt the Boer-Mulders
functions extracted from unpolarized $p+D$ Drell-Yan process and the
Collins functions extracted in Ref.~\cite{anselmino07} to calculate
the SIDIS $\cos 2 \phi$ asymmetries of charged pions measured at
ZEUS and JLab experiments. Explicitly, we will apply two sets of
Boer-Mulders functions. One is the set extracted in
Ref.~\cite{bing08}. The other is the new set we extract in this
paper from $p+D$ Drell-Yan data, by assuming the signs of
$h_1^{\p,u}$ and $h_1^{\p,d}$ to be different. Then we give
predictions for both $\cos 2 \phi$ asymmetries of the semi-inclusive
$\pi^+$ and $\pi^-$ production separately, which can be measured by
HERMES Collaboration (data to be analyzed) and the ongoing JLab
experiments.

\section{The $\cos 2 \phi$ asymmetry in unpolarized SIDIS}

\begin{figure}[t]
\begin{center}
\scalebox{0.65}{\includegraphics{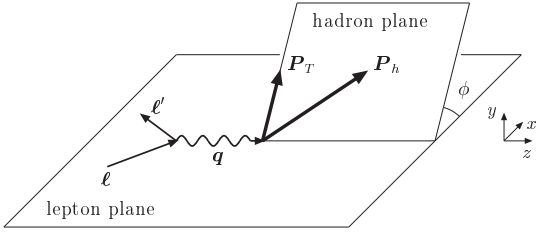}} \caption{\small Lepton
and hadron planes in semi-inclusive deep inelastic scattering.}
\label{plane}
\end{center}
\end{figure}

The process we are interested in reads:
\begin{equation}
l (\ell) \, + \, p (P) \, \rightarrow \, l' (\ell') \, + \, h (P_h)
\, + \, X (P_X)\,,\label{sidis}
\end{equation}
and the SIDIS cross section is expressed in terms of the invariants
\begin{equation}
x = \frac{Q^2}{2 \, P \cdot q}, \;\;\; y =  \frac{P \cdot q}{P \cdot
\ell} , \;\;\; z = \frac{P \cdot P_h}{P \cdot q}\,,
\end{equation}
where $ q = \ell - \ell'$ and $Q^2 \equiv - q^2$. We adopt a
reference frame such that the virtual photon and the target proton
or deuteron are collinear and directed along the $z$ axis, with the
photon moving in the positive $z$ direction . We denote by
$\Vec{k}_T$ the transverse momentum of the quark inside the proton,
and by $\Vec{P}_T$ the transverse momentum of the hadron $h$. The
transverse momentum of hadron $h$ with respect to the direction of
the fragmenting quark is $\Vec{p}_T$. All azimuthal angles are
referred to the lepton scattering plane and $\phi$  is the azimuthal
angle of the hadron $h$, as seen in Fig.~\ref{plane}.

Taking the intrinsic motion of quarks into account, in leading order
the azimuthal independent part of the SIDIS differential cross
section reads
\begin{eqnarray}
& &  \frac{\D \sigma}{\D x \, \D y \, \D z \, \D^2 \Vec P_T}
 =  \frac{2 \pi \alpha_{\rm em}^2 s}{Q^4} \, \sum_a
e_a^2 \, x [1 + (1 - y)^2] \nonumber \\
& &  \hspace{-1cm} \times \, \int \D^2 \Vec k_T \, \int \D^2 \Vec
p_T \, \delta^2 (\Vec P_T - z \Vec k_T - \Vec p_T) \, f_1^a (x, \Vec
k_T^2) \, D_1^a (z, \Vec p_T^2) \,, \nonumber \\ \label{cross1}
\end{eqnarray}
where $f_1^a (x, \Vec k_T^2)$ is the $\Vec k_T$-dependent
unpolarized distribution of quark with flavor $a$ and $D_1^a (z,
\Vec p_T^2)$ is the $\Vec k_T$-dependent fragmentation function of
quark. We recall that the non-collinear factorization theorem for
SIDIS has been proven by Ji, Ma and Yuan~\cite{Ji:2004xq} for $P_{T}
\ll Q$. We should remind that in above factorization formula an
additional soft factor should be taken into account. However in all
other phenomenological treatments this soft factor is neglected,
too.

According to the Cahn effect~\cite{cahn}, the non-collinear
transverse-momentum kinematics will generate a $\cos 2 \phi$
contribution to the unpolarized SIDIS cross section
\begin{eqnarray}
& & \left. \frac{\D \sigma^{(1)}}{\D x \, \D y \, \D z \, \D^2 \Vec
P_T} \right \vert_{\cos 2 \phi}
 =  \frac{8 \pi \alpha_{\rm em}^2 s}{Q^4} \, \sum_a
e_a^2 \, x (1 - y) \nonumber \\
& & \hspace{0cm} \times \,
 \int \D^2 \Vec k_T \, \int \D^2 \Vec p_T
\,\delta^2 (\Vec P_T - z \Vec k_T - \Vec p_T)
\nonumber \\
& & \hspace{0cm} \times \, \frac{2 \, (\Vec k_T \cdot \Vec h)^2 -
\Vec k_T^2}{Q^2} \, f_1^a (x, \Vec k_T^2) \, D_1^a (z, \Vec p_T^2)\,
\cos 2 \phi \,, \label{cross2}
\end{eqnarray}
 where $\Vec h \equiv \Vec P_T/P_T$. Notice that this contribution
is of order $k_T^2/Q^2$, so it is a kinematically higher twist
(twist-4) effect. Some caution are needed when implementing the Cahn
effect into the $\cos 2\phi$ asymmetries in unpolarized
cross-section, since the dynamical twist-4 contribution to SIDIS is
still unknown. Doubts about factorization for the twist-3 level were
mentioned in Refs.~\cite{Bacchetta:2008xw,Gamberg:2006ru}. Such
doubts will likely apply even more for twist-4. However we will not
discuss the detail of factorization at higher twist in this paper
and will apply Eq.~(\ref{cross2}) to calculate the $\cos 2 \phi$ at
the order of $k_T^2/Q^2$.

Another mechanism~\cite{bm} which can produce the $\cos 2 \phi$
asymmetry involves the coupling of $h_1^{\perp}$ and  the Collins
fragmentation function $H_1^{\perp}$, which is a leading twist
effect. The explicit expression of this contribution to the cross
section is
\begin{eqnarray}
& & \left. \frac{\D \sigma^{(2)}}{\D x \, \D y \, \D z \, \D^2 \Vec
P_T} \right \vert_{\cos 2 \phi}
 =  \frac{4 \pi \alpha_{\rm em}^2 s}{Q^4} \, \sum_a
e_a^2 \, x (1 - y) \nonumber \\
& & \hspace{-0.5cm} \times \,
 \int \D^2 \Vec k_T \, \int \D^2 \Vec p_T
\,\delta^2 (\Vec P_T - z \Vec k_T - \Vec p_T)
\nonumber \\
& & \hspace{-0.5cm} \times \, \frac{2 \, \Vec h \cdot \Vec k_T \,
\Vec h \cdot \Vec p_T - \Vec k_T \cdot \Vec p_T}{z  M M_h} \,
h_1^{\perp a} (x, \Vec k_T^2) \, H_1^{\perp a} (z, \Vec p_T^2)\,
\cos 2 \phi \,,\nonumber \\ \label{cross3}
\end{eqnarray}
where $M$ represents the mass of the target nucleon and $M_h$ the
mass of final-state produced hadron. It should be noticed that this
is a leading-twist contribution, not suppressed by inverse powers of
$Q$.

The $\cos 2 \phi$ asymmetry measured in experiments is defined as
\begin{equation}
\nu = \frac{\int \D \sigma \, \cos 2 \phi}{\int \D
\sigma}\,,\label{cross4}
\end{equation}
where the integrations are performed over the measured ranges of $x,
y, z$ and  $P_T$. Using Eqs.~(\ref{cross1}) to (\ref{cross4}), the
$\cos 2 \phi$ asymmetry for unpolarized SIDIS $\nu$ is given by
\begin{equation}
\nu= \frac{\int \sum_a e_a^2  2 x (1- y) \{ \frac{1}{2} \mathcal{B}
[h_1^{\perp a}, H_1^{\perp a}]+\mathcal{C} [f_1^a, D_1^a] \} }{\int
\sum_a e_a^2 x [1 + (1 - y)^2] \mathcal{A} [f_1^a,
D_1^a]},\label{asymmetry}
\end{equation}

where
\begin{equation}
\int \equiv \int_{P_T^{cut}}^{\infty}dP_T P_T
\int_{x_{min}}^{x_{max}} dx \int_{y_{min}}^{y_{max}} dy
\int_{z_{min}}^{z_{max}} dz
\end{equation}

and
\begin{eqnarray}
 \hspace{-5cm} \mathcal{A} [f_1^{a}, D_1^{a}] &\equiv& \int \D^2
\Vec k_T \, \int \D^2 \Vec p_T \, \delta^2 (\Vec P_T - z \Vec k_T -
\Vec p_T) \,
\nonumber \\
& & \hspace{-1cm}\times f_1^a (x, \Vec k_T^2) \, D_1^a (z, \Vec
p_T^2)
\nonumber \\
 & & \hspace{-2cm}= \int_0^{\infty} \D k_T \, k_T \, \int_0^{2 \pi} \D \chi \,
f_1^{a}(x, \Vec k_T^2) \, D_1^{a}(z, \vert \Vec P_T - z \Vec k_T
\vert^2)\,. \label{convol3}
\end{eqnarray}

\begin{eqnarray}
&&\mathcal{B} [h_1^{\perp a}, H_1^{\perp a}] \equiv
 \int \D^2 \Vec k_T \, \int \D^2 \Vec p_T
\,\delta^2 (\Vec P_T - z \Vec k_T - \Vec p_T)
\nonumber \\
  &\times & \, \frac{2 \, \Vec h \cdot \Vec k_T \, \Vec h \cdot \Vec
p_T - \Vec k_T \cdot \Vec p_T}{z  M M_h} \, h_1^{\perp a} (x, \Vec
k_T^2) \, H_1^{\perp a} (z, \Vec p_T^2)
\nonumber \\
& & \hspace{-1cm}= \int_0^{\infty} \D k_T  k_T  \int_0^{2 \pi} \D
\chi \, \frac{\Vec k_T^2 + (P_T/z)\, k_T \, \cos \chi - 2 \, \Vec
k_T^2 \, \cos^2 \chi}{M M_h}
\nonumber \\
 & & \hspace{-1cm}\times \, h_1^{\perp a}(x, \Vec k_T^2) \, H_1^{\perp a}(z, \vert
\Vec P_T - z \Vec k_T \vert^2)\,, \label{convol2}
\end{eqnarray}
\begin{eqnarray}
\mathcal{C} [f_1^{a}, D_1^{a}] &\equiv& \int \D^2 \Vec k_T \, \int
\D^2 \Vec p_T \,\delta^2 (\Vec P_T - z \Vec k_T - \Vec p_T)
\nonumber \\
& & \hspace{-0.5cm}  \times \, \frac{2 \, (\Vec k_T \cdot \Vec h)^2
- \Vec k_T^2}{Q^2} \, f_1^a (x, \Vec k_T^2) \, D_1^a (z, \Vec
p_T^2)\,
\nonumber \\
& & \hspace{-0.5cm}= \int_0^{\infty} \D k_T \, k_T \, \int_0^{2 \pi}
\D \chi \, \frac{2 \Vec k_T^2 \, \cos^2 \chi - \Vec k_T^2}{Q^2}
\nonumber \\
& &  \hspace{-0.5cm}\times \,  f_1^{a}(x, \Vec k_T^2) \, D_1^{a}(z,
\vert \Vec P_T - z \Vec k_T \vert^2)\,, \label{convol1}
\end{eqnarray}
with $\chi$  the angle between $\Vec P_T$ and $\Vec k_T$.

\section{Sets of the boer-mulders and collins functions used in our calculation}

In order to calculate the $\cos 2 \phi$ asymmetry given in the last
section, the forms of the $k_T$- and $p_T$-dependent distribution
and fragmentation functions appearing in Eqs.~(\ref{convol1}),
(\ref{convol2}) and (\ref{convol3}) should be provided.

Individual information of Boer-Mulders functions can be obtained
from the unpolarized $\pi + N$ Drell-Yan data~\cite{na10,conway}
which have been measured two decades ago, and most recently, the
unpolarized $p+D$ Drell-Yan data~\cite{e866} was measured by
E866/NuSea Collaboration. In Ref.~\cite{bing08}, based on E866/NuSea
data, we have extracted a set of $h_1^{\perp}(x, \Vec k_T^2)$ for
$u$, $d$, $\bar{u}$ and $\bar{d}$ quarks from the following form of
parameterizations:
\ba h_1^{\p,u}(x)&=&\omega H_u\,x^c\,(1-x)\,f_1^u(x),\label{p1new}\\
h_1^{\p,d}(x)&=&\omega H_d\,x^c\,(1-x)\,f_1^d(x),\label{p2new}\\
h_1^{\p,\bar{u}}(x)&=&\frac{1}{\omega}H_{\bar{u}}\,x^c\,(1-x)\,f_1^{\bar{u}}(x),\label{p3new}\\
h_1^{\p,\bar{d}}(x)&=&\frac{1}{\omega}H_{\bar{d}}\,x^c\,(1-x)\,f_1^{\bar{d}}(x),\label{p4new}
 \ea
where $\omega$ is a free coefficient, which can be determined in the
measurement of the unpolarized $p \bar{p}$ Drell-Yan process
discussed in Ref.~\cite{bing08}. The transverse momentum dependence
of the Boer-Mulders functions is expressed as
\begin{equation}
h_1^{\perp,q}(x,\mathbf{k}_T^2)=h_1^{\perp,q}(x)\frac{\textrm{exp}(-\mathbf{k}_{T}^{2}/
p_{bm}^2)}{\pi p_{bm}^2},\label{bmexp}
\end{equation}
in a Gaussian model with width $p_{bm}^2 $. The parameters extracted
in Ref.~\cite{bing08} are shown in the second column of
Table.~\ref{tab}, and is labeled as Set I.

\begin{table}[t]
\caption{\label{tab} Best fit of the Boer-Mulders functions
extracted from E866/NuSea $p + d$ Drell-Yan data. Set I is the
result in Ref.~\cite{bing08}, and Set II is the new result in this
work.} ~~ ~~~~~~~~~ ~~~~~~~~~

\begin{tabular}{|c|c|c|}
\hline
~          &  Set I  &   Set II \\
\hline
$H_u$           &  3.99 &   4.44     \\
\hline
$H_d$           &  3.83 &   -2.97  \\
\hline
$H_{\bar{u}}$           &  0.91&  4.68 \\
\hline
$H_{\bar{d}}$          &  -0.96& 4.98 \\
\hline
$p_{bm}^2$          & 0.161 &  0.165    \\
\hline
c                   &0.45 &   0.82  \\
\hline
$\chi^2/d.o.f.$      & 0.79 &  0.79 \\
\hline
\end{tabular}
\end{table}

\begin{figure}
\begin{center}
\scalebox{0.8}{\includegraphics[40pt,20pt][490pt,170pt]{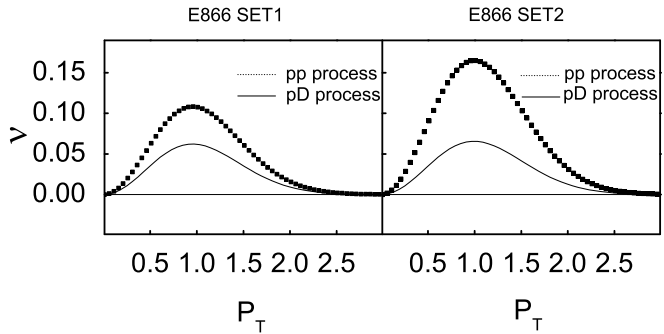}}
\caption{\small The $P_T$-dependent  $\cos 2 \phi$ asymmetries for
unpolarized $p+p$ and $p+D$ Drell-Yan processes of the FNAL
E866/NuSea Collaboration. The left part is taken of Set I , and the
right part is taken of Set II.}\label{e866pppd}
\end{center}
\end{figure}

The function $h_1^{\p}$ for $u$ and $d$ quarks given in Set I are of
the same sign and of similar size. It coincides with recent lattice
calculations ~\cite{lattice}, the approaches based on generalized
parton distributions (GPD) ~\cite{gpds,burkardt07}, the calculations
in Ref.~\cite{gamberg07}, and also a new quark-spectator-diquark
calculation~\cite{bct08}. However one can not exclude the
possibility that the functions $h_1^{\p,u}$ and $h_1^{\p,d}$ could
be of different signs, as the axial-diquark model calculation
predicts~\cite{bsy04}. Based on this assumption, we therefore
explicitly consider the case that the signs for $h_1^{\p,u}$ and
$h_1^{\p,d}$ are different, to extract another set of Boer-Mulders
functions. We give this result in Table.~\ref{tab}, labeled as Set
II. With both sets of $h_1^{\p}$, we then give predictions for the
$\cos 2 \phi$ asymmetry in unpolarized $p+p$ Drell-Yan process at
E866/NuSea, which are shown in Fig.~\ref{e866pppd}. As one can see,
the sizes of the asymmetries in $p+p$ Drell-Yan from Set I and Set
II are of 50\% difference. Thus the coming measurement in
unpolarized $p+p$ Drell-Yan by E866/NuSea Collaboration can
distinguish which set is preferred by data. In this work, we will
consider both sets of $h_1^{\p}$ to calculate the $\cos 2\phi$
asymmetry in SIDIS.

The possible values of $\omega$ should be constrained by a
positivity bound for Boer-Mulders functions:
 \ba \frac{k_{T}}{M}
h^\p_1(x,\mathbf{k}_{T}^2)  \le  f_1(x,\mathbf{k}_{T}^2).
 \ea
Therefore for the case of Set I, we can get a constraint for
$\omega$ as $0.177 \leq \omega \leq 1.330$ for all $x$ and $k_T$, in
order to obey the positive bound. For the case of  Set II, the
constraint for $\omega$ is $0.490 \leq \omega \leq 1.670$.

The transverse momentum dependence of the distribution function
$f_1(x,\mathbf{k}_T^2)$ and fragmentation function
$D_1(z,\mathbf{p}_T^2)$ is also given in the Gaussian form adopted
in \cite{anselmino07}:
 \be f_1^q(x,\mathbf{k}_T^2)=f_1^q(x)\,
\frac{\textrm{exp}\,\, (-\mathbf{k}_T^2/p_{unp}^2)}{\pi \, p_{unp}^2
}, \label{unp} \ee
\be
D_1^{\perp,q}(z,\mathbf{p}_T^2)=D_1^{\perp,q}(z)\frac{\textrm{exp}(-\mathbf{p}_{T}^{2}/
p_{f}^2)}{\pi p_{f}^2}.\label{D1}  \ee

For the integrated unpolarized distribution function $f_1^q(x)$ we
adopt the MRST2001 (LO set) parametrization ~\cite{mrst}, and for
the unpolarized fragmentation function $D_1^q(z)$, we adopt the
parametrization given by Kretzer~\cite{kretzer}. We take the
Gaussian width $ p_{unp}^2 = 0.25$ and $ p_{f}^2 = 0.2$, following
the choice in Ref.~\cite{punp}, which was obtained by fitting the
azimuthal dependence of the unpolarized SIDIS cross section.

Concerning the Collins functions, independent information about them
can be obtained from the measurement on the angular dependence of
hadron pair production in $e^+ e^- \to h^+ h^- +X$ process through
$H_1^{\p,q} \times H_1^{\p,\bar{q}}$.  Experimental data on $e^+
e^-$ has been obtained by Delphi at CERN a decade ago and recently
by Belle~\cite{belle06} at KEK. Access to the Collins function is
also possible through the measurement of Collins single-spin
asymmetry in transversely polarized SIDIS process, which has been
measured by HERMES~\cite{Air2004tw},
COMPASS~\cite{compass,compass07} and JLab in recently years. Several
groups have extracted the Collins function from the those $e^+ e^-$
and the SIDIS data. In Ref.~\cite{vy05} a $\Vec k_T^2$-1/2 moment of
Collins function $H_1^{\p,1/2}(z)$ was introduced and parameterized
to fit the HEREMES SIDIS data. The authors of Ref.~\cite{egp06}
adopted a Gaussian form for the $\Vec k_T$ dependence of
$H_1^\p(z,\Vec k_T^2)$ to extract the Collins functions from $e^+
e^-$ data of Belle and SIDIS data of HERMES. In
Ref.~\cite{anselmino07} a different parametrization for Collins
function is adopted to perform a global analysis based on the $e^+
e^-$ data at Belle and the SIDIS data at HERMES and COMPASS. All the
extractions apply the concept of equally sizable favored and
unfavored fragmentation function in order to describe the
experimental data successfully. In this paper we adopt the
parameterizations given in Ref.~\cite{anselmino07}, which have the
form
 \be
 H_1^{\perp}(z,\mathbf{p}_T^2) = \frac{z m_h }{\mathbf{p}_T}
\mathcal {N}_q^C(z) D_1(z) h(\mathbf{p}_T)
\frac{\textrm{exp}(-\mathbf{p}_{T}^{2}/ p_{f}^2)}{\pi p_{f}^2},
\label{coll-funct}
 \ee
with \ba \mathcal {N}_q^C(z)=\mathcal {N}_q^C
z^{\gamma}(1-z)^{\delta}\frac{(\gamma+\delta)^{(\gamma+\delta)}}{\gamma^{\gamma}\delta^{\delta}},
\ea \ba h(\mathbf{p}_T)=(2e)^{1/2}\frac{\mathbf{p}_T}{M
}e^{-\frac{\mathbf{p}_T^2}{M^2 }},
 \ea
with the parameters as follows
\begin{eqnarray}
\mathrm{Set~I} : \mathcal {N}_{fav}^C=0.35, \, \mathcal
{N}_{unf}^C=-0.85, \nonumber\\
\gamma=1.14, \, \delta=0.14, \, M^2=0.70,
\nonumber\\
\mathrm{Set~II} : \mathcal {N}_{fav}^C=0.41, \, \mathcal
{N}_{unf}^C=-0.99,\nonumber\\
\gamma=0.81, \, \delta=0.02, \, M^2=0.88.
\end{eqnarray}

In all the numerical calculations given below, we will apply the
parametrization of Collins functions given in Set I. The results
from Set II are very similar.

Finally we should emphasize that the the signs of T-odd distribution
functions should be reversed~\cite{collins02} from Drell-Yan process
to DIS process, by the presence of the path-ordered exponential
(Wilson line) in the gauge-invariant definition of the transverse
momentum dependent parton distributions. Therefore the sets of
Boer-Mulders functions shown in Table.~\ref{tab} should be reversed
the signs when they are applied in the calculation of SIDIS process.

\section{Results and predictions}

First, we will calculate the $\cos 2 \phi$ asymmetries in SIDIS
under the kinematics of ZEUS, and compare our prediction with their
data~\cite{zeus2006}. ZEUS employs an unpolarized positron beam with
energy 27.6 GeV to collide with a proton beam of 820 $\textrm{GeV}$.
The fragmenting hadrons are detected with pseudorapidity
$\eta^{HCM}=\frac{1}{2}\textrm{ln}\frac{x}{y}$ and minimum
transverse energy $E_{T,min}^{HCM}$. Here
$x=\sqrt{\frac{Q^2+Q_T^2}{s}}\,e^{ \eta^{HCM}}$ and
$y=\sqrt{\frac{Q^2+Q_T^2}{s}}\,e^{ -\eta^{HCM}}$. The $\cos 2 \phi$
asymmetries of charged hadrons are measured, most of which are
charged pions. Therefore we will calculate the $\cos 2 \phi$
asymmetries of charged pions as an approximation to the $\cos 2
\phi$ asymmetries of charged hadrons. Recent HERMES
data~\cite{Diefenthaler:2006vn} indicate that the Sivers and Collins
asymmetries of  production are also sizable. Therefore one can
expect that the $\cos 2 \phi$ asymmetries of $K^\pm$ production
could be not so small. However the contribution of $K^\pm$ to the
$\cos 2 \phi$ asymmetry of charged hadron production will be minor
comparing to that of pions, as the production rate of $K^\pm$ is
much smaller than that of pions.

The ZEUS kinematics is characterized by the following:
\begin{eqnarray}
&&0.01<x<0.1, ~~  0.2<y<0.8,\nonumber\\
&&100<Q^2<8000 \textrm{GeV}^2, ~~ p_T>0.15\textrm{GeV}.\
\end{eqnarray}
Therefore the ZEUS experiment was taken at small $x$ and very high
$Q^2$, with average values of $Q^2$ around 750 $\textrm{GeV}^2$. In
this kinematics regime the perturbative
contribution~\cite{georgi,mendez,konig,chay} from gluon emission and
splitting in NLO QCD is highly relevant. In Ref.~\cite{ma08} this
contribution to the $\cos 2 \phi$ asymmetry in SIDIS has been
calculated showing that it is substantial  at ZEUS kinematics. In
this paper, the contribution by the Boer-Mulders functions as well
as the Cahn effect to the $\cos 2 \phi$ asymmetries is primarily
devoted to making predictions for the low-$Q^2$ (of few
$\textrm{GeV}^2$) and low $P_T$~\cite{boer2001} regimes, where gluon
emission is quite irrelevant. Therefore in the entire paper, we will
not consider perturbative contribution in our calculation.

\begin{figure}
\begin{center}
\scalebox{0.85}{\includegraphics[40pt,30pt][500pt,267pt]{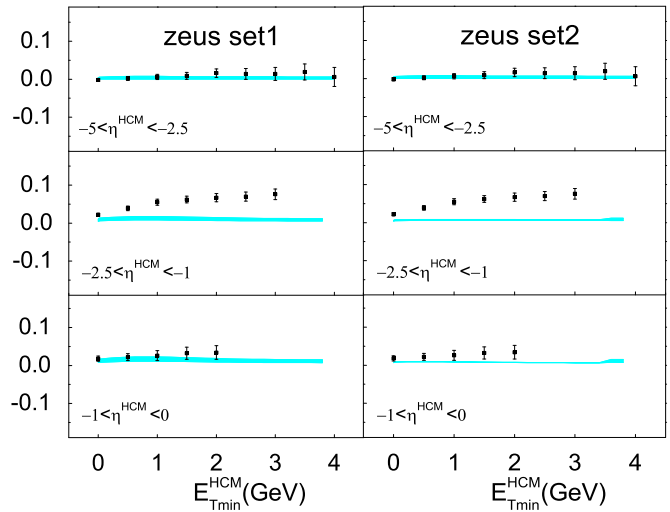}}
\caption{\small The $E_{T,min}^{HCM}$-dependent $\cos 2 \phi$
asymmetries at ZEUS with pseudorapidity $-5 < \eta^{HCM} \leq -2.5$,
$-2.5 < \eta^{HCM} \leq -1$ and $-1 < \eta^{HCM} \leq 0$. The left
column is from Boer-Mulders functions of Set I with $0.2 \leq \omega
\leq 0.5$, the right column is from Boer-Mulders functions of Set II
with $0.5 \leq \omega \leq 0.8$.}\label{zeus2006e}
\end{center}
\end{figure}

\begin{figure}
\begin{center}
\scalebox{0.8}{\includegraphics[40pt,20pt][490pt,170pt]{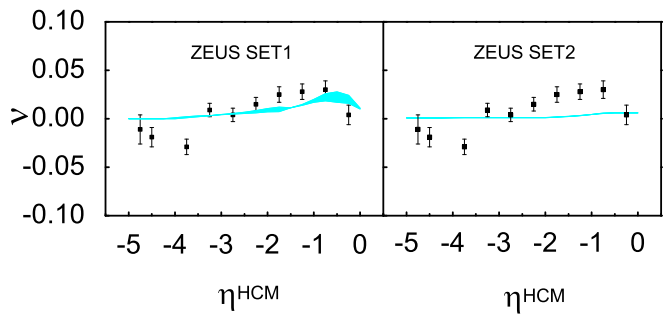}}
\caption{\small The $\eta^{HCM}$-dependent $\cos 2 \phi$ asymmetries
at ZEUS. The left column is from Boer-Mulders functions of Set I
with $0.2 \leq \omega \leq 0.5$, the right column is from
Boer-Mulders functions of Set II with $0.5 \leq \omega \leq
0.8$.}\label{zeus2006nu}
\end{center}
\end{figure}

For the case of Set I, considering the constraint for $\omega$ value
by the positivity bound for Boer-Mulders functions, with $\omega$
from 0.2 to 0.5 which is a reasonable range, we give our prediction
for the $\cos 2 \phi$ asymmetries as function of the minimum
transverse energy $E_{T,min}^{HCM}$ and $\eta^{HCM}$, as shown in
the left sides of Figs.~\ref{zeus2006e} and \ref{zeus2006nu},
respectively. In the same figures, we also give prediction from
Boer-Mulders functions of Set II, with $\omega$ from 0.5 to 0.8.

\begin{figure}
\begin{center}
\scalebox{0.8}{\includegraphics[40pt,20pt][490pt,175pt]{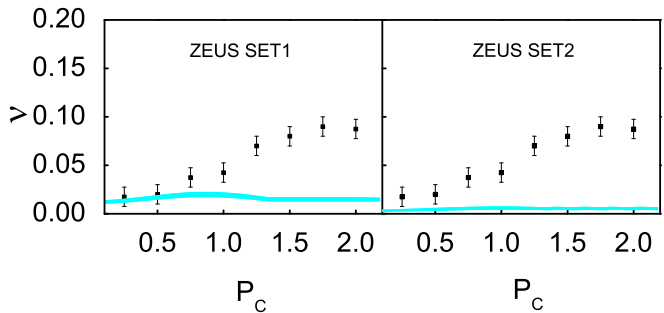}}
\caption{\small The $P_c$-dependent $\cos 2 \phi$ asymmetries at the
earlier ZEUS experiment. Data are from Ref.~\cite{zeus2000}. The
left column is from Boer-Mulders functions of Set I with $0.2 \leq
\omega \leq 0.5$, the right column is from Boer-Mulders functions of
Set II with $0.5 \leq \omega \leq 0.8$.}\label{zeus2000pc}
\end{center}
\end{figure}

\begin{figure}
\begin{center}
\scalebox{0.8}{\includegraphics[40pt,20pt][490pt,170pt]{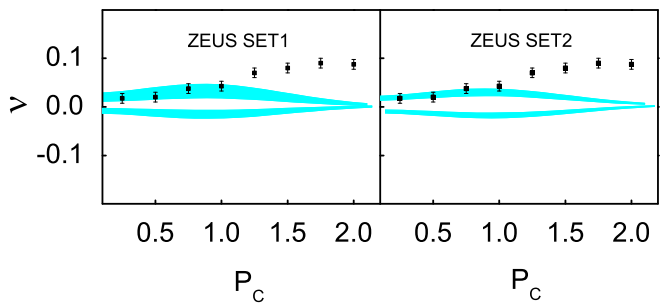}}
\caption{\small The $P_c$-dependent $\cos 2 \phi$ asymmetries for
$\pi^+$ and $\pi^-$ separately compared with the earlier ZEUS
experimental data. The left column is from Boer-Mulders functions of
Set I with $0.2 \leq \omega \leq 0.5$, the right column is from
Boer-Mulders functions of Set II with $0.5 \leq \omega \leq
0.8$.}\label{zeus2000pcseparate}
\end{center}
\end{figure}

The earlier ZEUS experiment~\cite{zeus2000} measured the $\cos 2
\phi$ asymmetries as the function of transverse momentum cutoff
$P_c$. In Fig.~\ref{zeus2000pc} we also give our prediction for this
experiment with both sets of Boer-Mulders functions. Furthermore, we
give the predictions for the $\cos 2 \phi$ asymmetries of $\pi^+$
and $\pi^-$ production vs $P_c$ separately in
Fig.~\ref{zeus2000pcseparate}. As shown in Figs.~(\ref{zeus2000pc})
and (\ref{zeus2000pcseparate}), our results are only reliable for
low values $P_c$. At larger $P_c$ our result underestimates the
asymmetry. The reason is that we have not included perturbative
contribution which gives the main contribution at high-$P_T$. In
Ref.~\cite{ma08} the perturbative contribution for ZEUS has been
calculated and an agreement between the theoretical calculation and
data at large $P_c$ was found.

The $\cos 2 \phi$ asymmetries were also measured at CERN by
EMC~\cite{Arneodo:1986cf}, but with low precision. The $\cos 2 \phi$
asymmetries measured at an earlier ZEUS experiment~\cite{zeus2000}
and at EMC~\cite{Arneodo:1986cf}, were estimated in Ref.~\cite{lu06}
with a $u$-quark dominating model for $h_1^\perp$ and Gaussian
ansatz for the Collins function. In the case of Set I, our results
for the the $\cos 2 \phi$ asymmetries as a function of the $P_T$
cutoff $P_c$ up to 0.5 GeV agree with the predictions of
Ref.~\cite{lu06}.

The experiment in Jefferson Lab (JLab) also measured~\cite{jlab07}
the $\cos 2 \phi$ asymmetries of the semi-inclusive charged pion
production as a function of $x$ and $z$, from both proton and
deuteron targets, using an electron beam with energy 5.5 GeV
\cite{jlab07}. The kinematics at JLab is characterized by the
following ranges:
\begin{eqnarray}
&&0.2<x<0.5, ~ 0.4<y<0.9, ~ 0.3<z<1\nonumber\\
&&2<Q^2<4 \textrm{GeV}^2, ~ P_t^2<0.2 \textrm{GeV}^2.
\end{eqnarray}

\begin{figure}
\begin{center}
\scalebox{0.78}{\includegraphics[35pt,20pt][450pt,182pt]{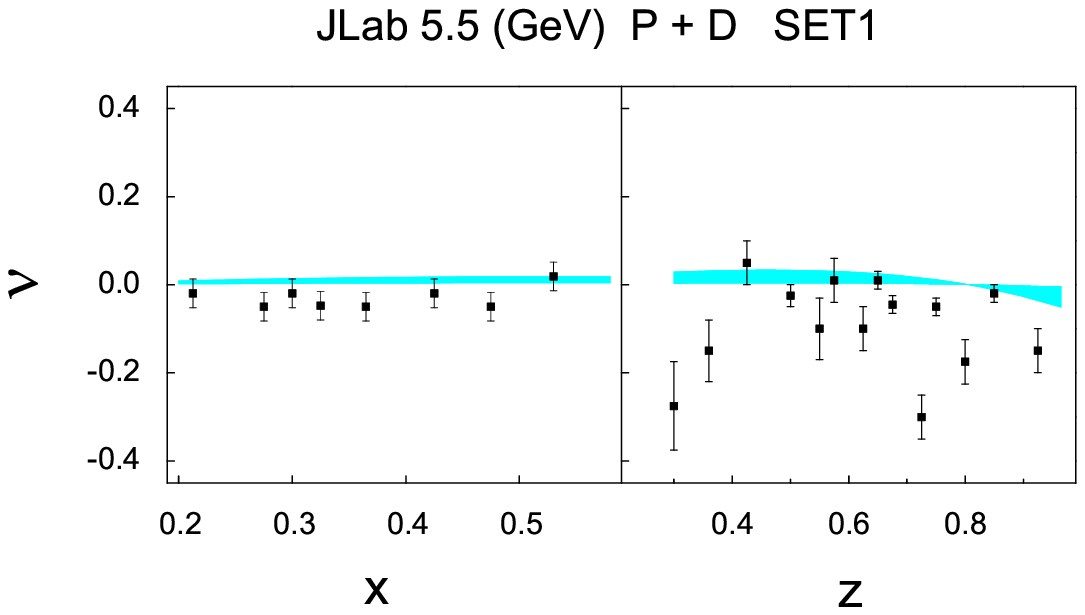}}
\caption{\small The $x$- and $z$-dependent $\cos 2 \phi$ asymmetries
at JLab with beam energy 5.5 GeV. In the calculation we apply
Boer-Mulders functions in Set I and take $0.2 \leq \omega \leq 0.5$.
Data are from Ref.~\cite{jlab07}}\label{set1jlab6pd}
\end{center}
\end{figure}

\begin{figure}
\begin{center}
\scalebox{0.78}{\includegraphics[35pt,20pt][450pt,182pt]{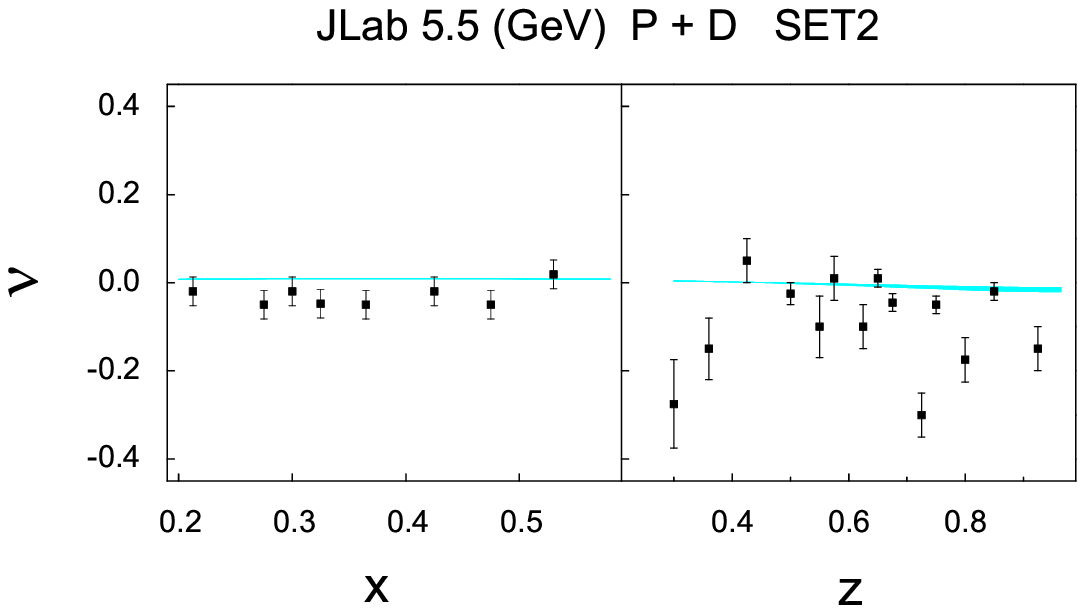}}
\caption{\small Same as Fig.~\ref{set1jlab6pd}, but from
Boer-Mulders functions in Set II. In the calculation we take $0.5
\leq \omega \leq 0.8$.}\label{set2jlab6pd}
\end{center}
\end{figure}

In Ref.~\cite{jlab07}, the asymmetries for $\pi^\pm$ on proton or
deuteron targets are combined together. Here we calculate
$x$-dependent and $z$-dependent $\cos 2 \phi$ asymmetry at JLab in
the same way, and the results are shown in Fig.~\ref{set1jlab6pd}
and Fig.~\ref{set2jlab6pd}, respectively. All the results are
obtained with the range of $\omega$ from 0.2 to 0.5 for Set I and
$\omega$ from 0.5 to 0.8 for Set II. In the calculation for a
deuteron target, we have used the isospin relation:
\begin{eqnarray}
f^{u/D} \approx f^{u/p}+f^{u/n}=f^u+f^d,
\\
f^{d/D} \approx f^{d/p}+f^{d/n}=f^d+f^u,
\\
f^{\bar{u}/D} \approx
f^{\bar{u}/p}+f^{\bar{u}/n}=f^{\bar{u}}+f^{\bar{d}},
\\
f^{\bar{d}/D} \approx f^{\bar{d}/p}+f^{
\bar{d}/n}=f^{\bar{d}}+f^{\bar{u}},
\end{eqnarray}
for both $f_1$ and $h_1^\p$.

\begin{figure}
\begin{center}
\scalebox{0.72}{\includegraphics[10pt,20pt][488pt,220pt]{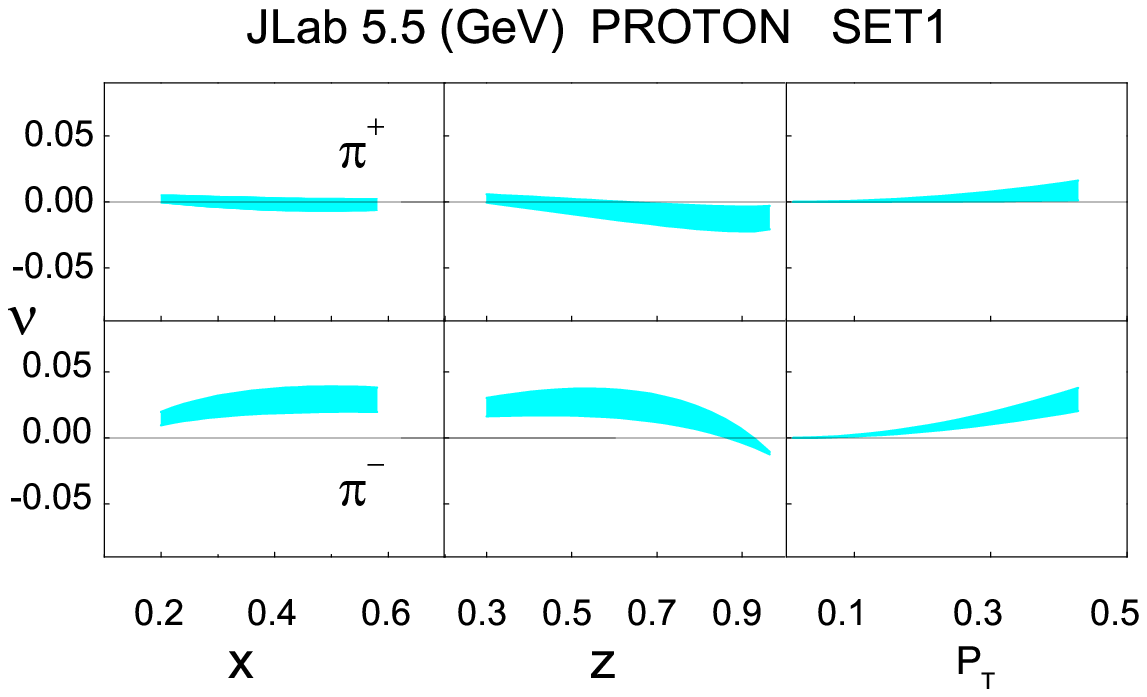}}
\caption{\small The $x$-dependent $\cos 2 \phi$ asymmetries at JLab
with 6 GeV on proton target. In the calculation we apply
Boer-Mulders functions in Set I and take $0.2 \leq \omega \leq
0.5$.}\label{set1jlab6p}
\end{center}
\end{figure}

\begin{figure}
\begin{center}
\scalebox{0.72}{\includegraphics[10pt,20pt][438pt,220pt]{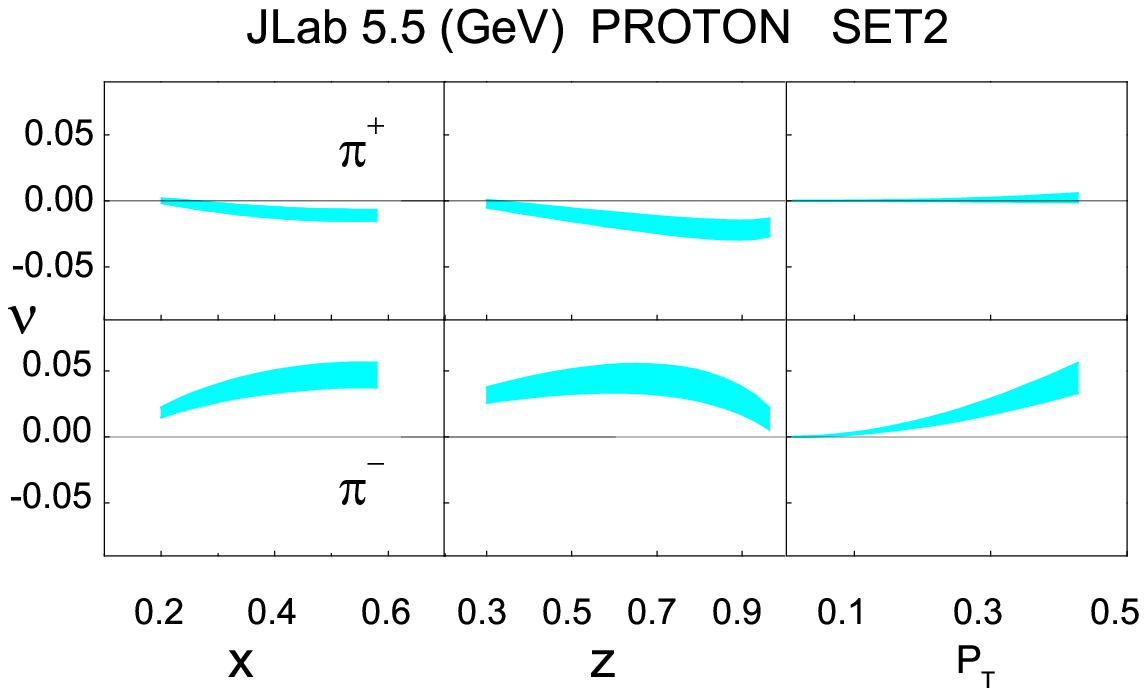}}
\caption{\small Same as Fig.~\ref{set1jlab6p}, but from Boer-Mulders
functions in Set II. In the calculation we take $0.5 \leq \omega
\leq 0.8$.}\label{set2jlab6p}
\end{center}
\end{figure}

\begin{figure}
\begin{center}
\scalebox{0.72}{\includegraphics[10pt,20pt][488pt,220pt]{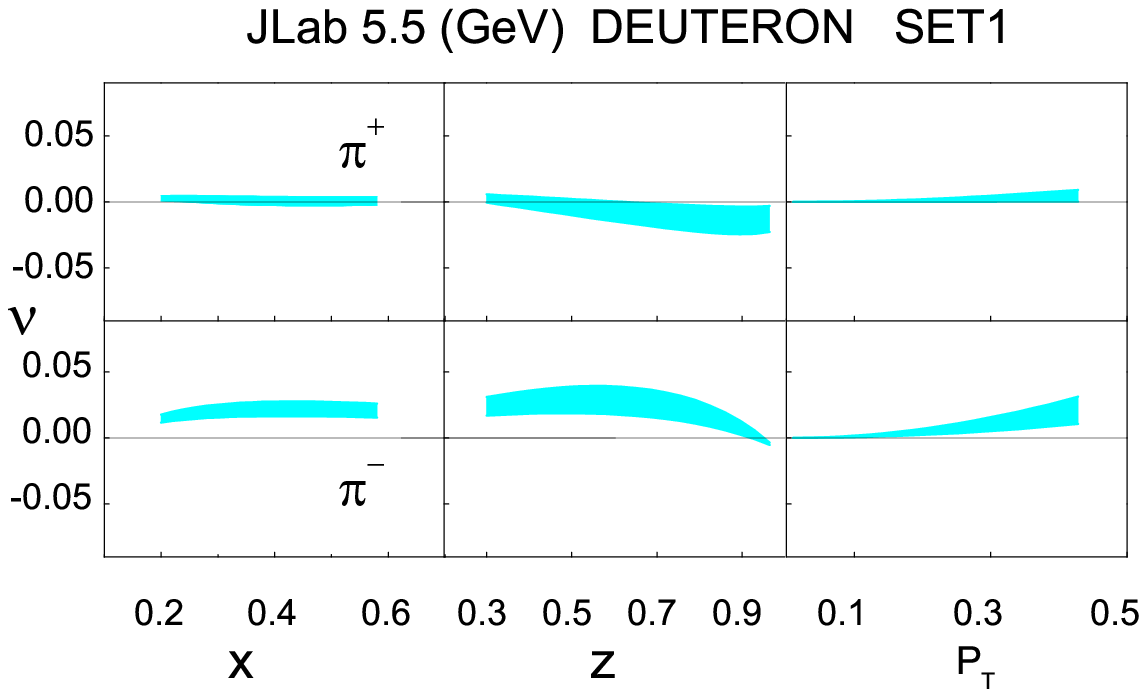}}
\caption{\small The $x$-dependent $\cos 2 \phi$ asymmetries at JLab
with 6 GeV on deuteron target. In the calculation we apply
Boer-Mulders functions in Set I and take $0.2 \leq \omega \leq
0.5$.}\label{set1jlab6d}
\end{center}
\end{figure}

\begin{figure}
\begin{center}
\scalebox{0.72}{\includegraphics[14pt,20pt][438pt,198pt]{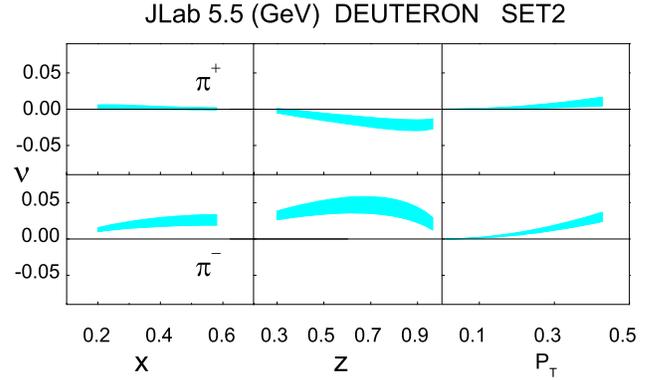}}
\caption{\small Same as Fig.~\ref{set1jlab6d}, but from Boer-Mulders
functions in Set II. In the calculation we take $0.5 \leq \omega
\leq 0.8$.}\label{set2jlab6d}
\end{center}
\end{figure}

Our predictions for ZEUS and JLab show that the $\cos 2 \phi$
asymmetries of charged hadrons (pions) are rather small. This is due
to the that the favored Collins functions adopted here are positive
while the unfavored ones are negative. Therefore when the
contribution of $\pi^+$ and $\pi^-$ production are combined
together, there is a cancelation which leads a small value.

Thus we suggest to measure and analyze the asymmetries of $\pi^+$
and $\pi^-$ production separately, where larger asymmetries may be
obtained. In Fig.~\ref{set1jlab6p} and Fig.~\ref{set1jlab6d} for the
case of Set I, we predict the $\cos 2 \phi$ asymmetries as the
functions of $x$ , $z$ and $P_T$, of both $\pi^+$ and $\pi^-$
production on proton target and deuteron target respectively, with a
5.5 GeV beam at JLab, showing a larger asymmetry for $\pi^-$
production than that of $\pi^+$ production. In the case of Set II
(Fig.~\ref{set2jlab6p} and Fig.~\ref{set2jlab6d}), we obtain a
larger asymmetry for the $\pi^-$ production and a smaller asymmetry
for the $\pi^+$ production. Therefore we expect the separate analyse
of $\cos 2 \phi$ asymmetries on $\pi^+$ and $\pi^-$ production of
JLab ~\cite{jlab07} will make us know more information about the
Boer-Mulders functions, especially for distinguishing which sets are
better to describe data.

\begin{figure}
\begin{center}
\scalebox{0.72}{\includegraphics[14pt,20pt][338pt,220pt]{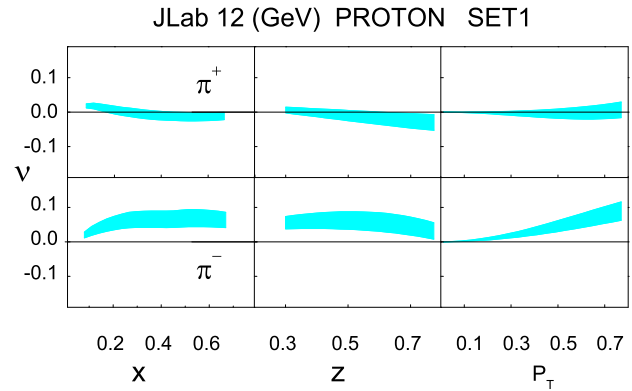}}
\caption{\small The $x$- , $z$- and $P_T$-dependent $\cos 2 \phi$
asymmetries at JLab with 12 GeV on proton target. In the calculation
we apply Boer-Mulders functions of Set I and take $0.2 \leq \omega
\leq 0.5$.}\label{set1jlab12p}
\end{center}
\end{figure}

\begin{figure}
\begin{center}
\scalebox{0.72}{\includegraphics[14pt,20pt][338pt,220pt]{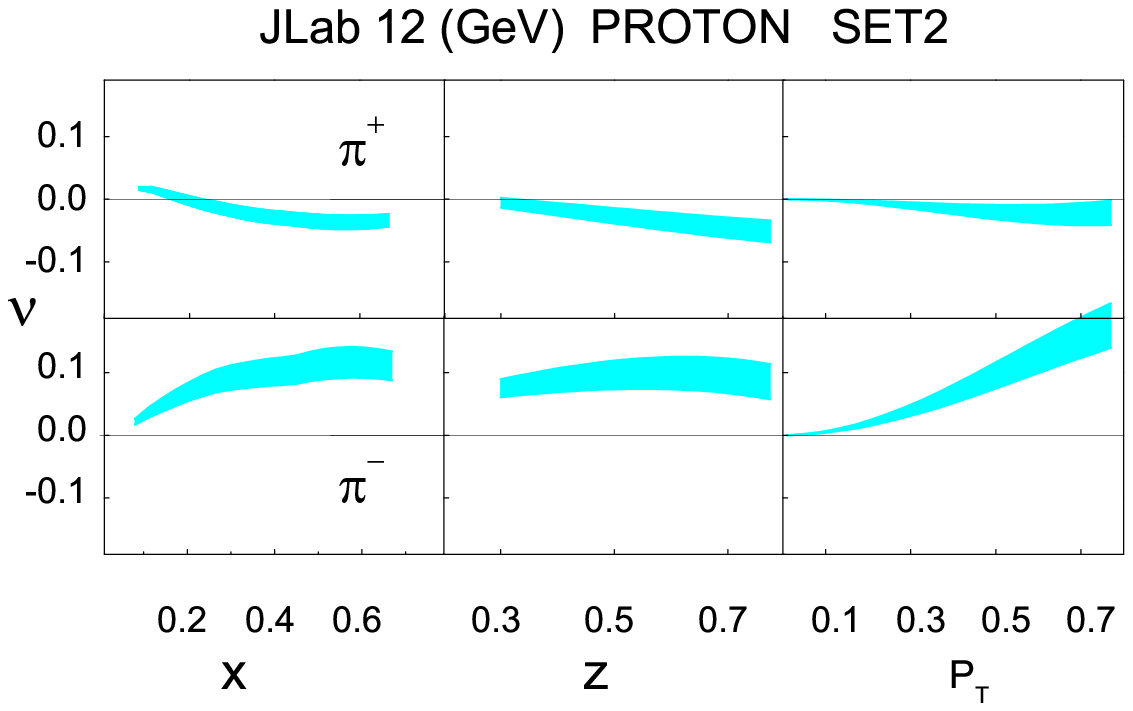}}
\caption{\small Same as Fig.~\ref{set1jlab12p}, but from
Boer-Mulders functions of Set II. In the calculation we take $0.5
\leq \omega \leq 0.8$.}\label{set2jlab12p}
\end{center}
\end{figure}

\begin{figure}
\begin{center}
\scalebox{0.72}{\includegraphics[14pt,20pt][338pt,220pt]{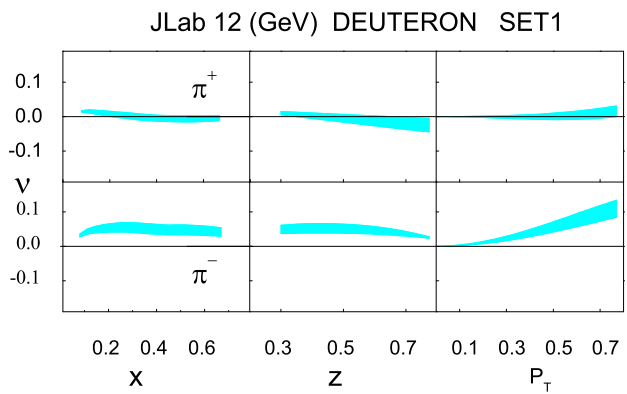}}
\caption{\small The $x$- , $z$- and $P_T$-dependent $\cos 2 \phi$
asymmetries at JLab with 12 GeV on deuteron target. In the
calculation we take $0.2 \leq \omega \leq 0.5$.}\label{set1jlab12d}
\end{center}
\end{figure}

\begin{figure}
\begin{center}
\scalebox{0.72}{\includegraphics[14pt,20pt][338pt,220pt]{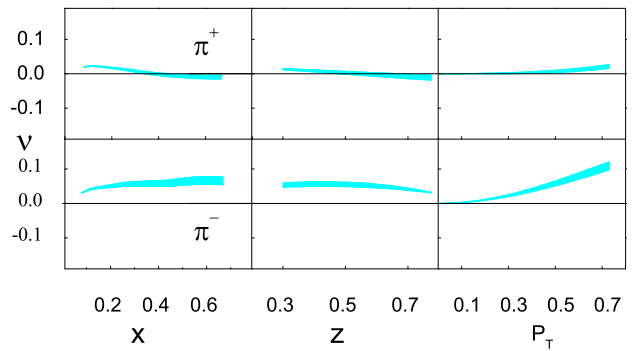}}
\caption{\small Same as Fig.~\ref{set1jlab12d}, but from
Boer-Mulders functions of Set II. In the calculation we take $0.5
\leq \omega \leq 0.8$.}\label{set2jlab12d}
\end{center}
\end{figure}

By applying the two sets of Boer-Mulders functions, we also give the
prediction on the $\cos 2 \phi$ asymmetries of semi-inclusive pion
production with electron beam energy of 12 GeV both for proton and
deuteron targets, which are applicable at JLab after the upgrade of
the beam energy. The asymmetries as the functions of  $x$ , $z$ and
$P_T$ are calculated with the kinematical regime
\begin{eqnarray}
&&0.08<x<0.7, ~~ 0.2<y<0.9, ~~ 0.3<z<0.8, \nonumber\\
&&1<Q<3 \textrm{GeV}, ~~ 1<E_{\pi}<9  ~~ \textrm{GeV},
\end{eqnarray}
and are shown in Fig.~\ref{set1jlab12p}, Fig.~\ref{set1jlab12d},
Fig.~\ref{set2jlab12p} and Fig.~\ref{set2jlab12d}.

\begin{figure}
\begin{center}
\scalebox{0.72}{\includegraphics[14pt,20pt][338pt,220pt]{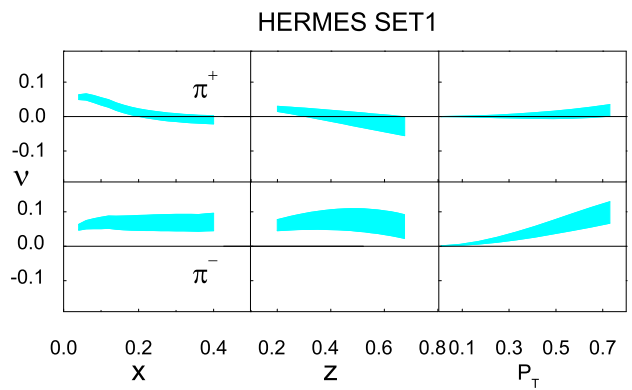}}
\caption{\small The $x$-, $z$-  and $P_T$-dependent $\cos 2 \phi$
asymmetries for proton target at HERMES. In the calculation we apply
Boer-Mulders functions of Set I and take $0.2 \leq \omega \leq
0.5$.}\label{set1hermes}
\end{center}
\end{figure}

\begin{figure}
\begin{center}
\scalebox{0.72}{\includegraphics[14pt,20pt][338pt,220pt]{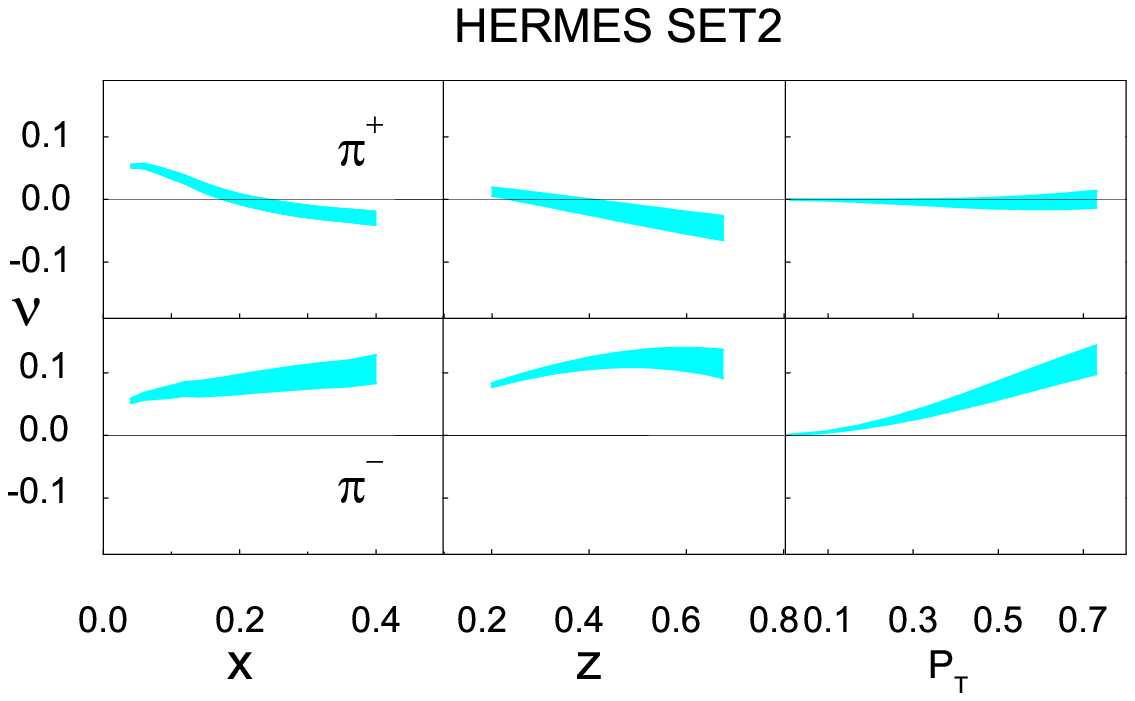}}
\caption{\small Same as Fig.~\ref{set1hermes}, but from Boer-Mulders
functions of Set II. In the calculation we take $0.5 \leq \omega
\leq 0.8$.}\label{set2hermes}
\end{center}
\end{figure}

Finally, we give the prediction on the $\cos 2 \phi$ asymmetries of
$\pi^+$ and $\pi^-$ production in SIDIS at HERMES which employs an
electron beam with energy 27.5 GeV off the proton target. From the
kinematics regime of HERMES:
\begin{eqnarray}
&&0.2<x<0.42, ~~ 0.2<y<0.8 ~~ 0.2<z<0.7, \nonumber\\
&&1<Q<3.87 \textrm{GeV} ~~ 4.5<E_{\pi}<13.5 \, \textrm{GeV},
\nonumber\
\end{eqnarray}
we calculate the $x$-, $z$-  and $P_T$-dependent $\cos 2 \phi$
asymmetries respectively from Boer-Mulders functions of Sets I and
II, as shown in Fig.~\ref{set1hermes} and Fig.~\ref{set2hermes},
respectively. We expect to know more information on the Boer-Mulders
functions with our prediction comparing with the data to be analyzed
in HERMES.

  From our prediction for forthcoming JLab and the HERMES
experiments, we arrive at a conclusion that the size of $\cos 2
\phi$ asymmetries in semi-inclusive $\pi^-$ production are somewhat
larger than that in $\pi^+$ production for the case of Set I, and
the $\cos 2 \phi$ asymmetries of semi-inclusive $\pi^-$ production
are larger than that of $\pi^+$ production for the case of Set II.
On the calculations of the $\cos 2 \phi$ asymmetries of
semi-inclusive $\pi^+$ production, the negative result from
Boer-Mulders functions and positive result from Cahn effect combine
to yield a very small $\pi^+$ asymmetry, while on the calculations
of the $\cos 2 \phi$ asymmetries of semi-inclusive $\pi^-$
production, the positive result from Boer-Mulders functions and
positive result from Cahn effect combine to yield a several percent
$\pi^-$ asymmetry. These results are also predicted in
Ref.~\cite{gamberg07} and in Ref.~\cite{ma08}. We suggest to measure
and analyze the $\cos 2 \phi$ asymmetries of semi-inclusive $\pi^+$
production and that of $\pi^-$ production separately. This will help
us to know more details of the Boer-Mulders functions as well as the
Collins functions.

\section{Conclusion}

In summary, we use the Boer-Mulders functions extracted from
unpolarized $p+D$ Drell-Yan process to study the $\cos 2 \phi$
asymmetries in SIDIS processes. Two sets of Boer-Mulders functions
are applied. The first set (Set I) is the result extracted in
Ref.~\cite{bing08}, while the second set (Set II) is the new result
presented in this work by explicitly assuming that the signs of
$h^{\p, u}$ and $h^{\p, d}$ are different. The predictions for the
$\cos 2 \phi$ asymmetry in unpolarized $p+p$ Drell-Yan from two sets
of Boer-Mulders function defer by 50\%, indicating that the
measurement in $p+p$ Drell-Yan can distinguish which set is
preferred by data. With both sets of $h^{\p}$, we estimate the $\cos
2 \phi$ asymmetries of charged pion production measured at ZEUS, and
at JLab with 5.5 GeV beam energy, respectively. Then we give
predictions for the $\cos 2 \phi$ asymmetries of $\pi^+$ and $\pi^-$
production at HERMES and at 12 GeV JLab experiments. We suggest that
measuring the $\cos 2 \phi$ asymmetries of $\pi^+$ production and
$\pi^-$ production separately will help to provide further
information on Boer-Mulders functions.


{\bf Acknowledgements.} This work is partially supported by National
Natural Science Foundation of China (Nos.~10721063, 10575003,
10505001, 10528510), by the Key Grant Project of Chinese Ministry of
Education (No.~305001), by the Research Fund for the Doctoral
Program of Higher Education (China), by Project of PBCT No. ACT-028
(Chile).

\end{document}